# Solid-state single-photon sources: recent advances for novel quantum materials

Martin Esmann[1,*], Stephen C. Wein[2,*], Carlos Antón-Solanas[3,*]

[1]Institute of Physics, Carl von Ossietzky University Oldenburg, 26129 Oldenburg, Germany

[2]Institute for Quantum Science and Technology and Department of Physics and Astronomy, University of Calgary, Calgary, Alberta, Canada

[3]Depto. Física de Materiales, Instituto Nicolás Cabrera, Instituto de Física de la Materia Condensada, Universidad Autónoma de Madrid, 28049 Madrid, Spain

*Email address: m.esmann@uni-oldenburg.de, wein.stephen@gmail.com, carlos.anton@uam.es

**Abstract**

In this review, we describe the current landscape of emergent quantum materials for quantum photonic applications. We focus on three specific solid-state platforms: single emitters in monolayers of transition metal dichalcogenides, defects in hexagonal boron nitride, and colloidal quantum dots in perovskites. These platforms share a unique technological accessibility, enabling the rapid implementation of testbed quantum applications, all while being on the verge of becoming technologically mature enough for a first generation of real-world quantum applications.

The review begins with a comprehensive overview of the current state-of-the-art for relevant single-photon sources in the solid-state, introducing the most important performance criteria and experimental characterization techniques along the way. We then benchmark progress for each of the three novel materials against more established (yet complex) platforms, highlighting performance, material-specific advantages, and giving an outlook on quantum applications. This review will thus provide the reader with a snapshot on latest developments in the fast-paced field of emergent single-photon sources in the solid-state, including all the required concepts and experiments relevant to this technology.

## 1. Introduction

In the current technological landscape, **quantum photonics** has given rise to a transformative field with impactful applications in many domains [1]. A significant part of quantum optical technologies harnesses the properties of quantum materials to manipulate light and matter at the nanoscale. These advancements offer a wide range of applications ranging from quantum-secure communication networks [2–4] to more efficient computer algorithms [5,6]; these breakthroughs hold the potential to address challenges in cryptography, computing power, sensing and metrology, medical and biological applications, and materials science.

This review focuses on one specific building block of quantum photonics: **solid-state sources of single photons**, more specifically on **emergent quantum materials** for their realization. The efficient generation of single photons has ubiquitous relevance in photonic applications [7] related to sensing [8], communication [9], simulation [10], and computation [11,12]. In this sense, research on new quantum materials may enable the application-tailoring of **advanced properties**. For example, it is desirable to develop sources with an emission wavelength in the telecommunication C (1550nm) or O (1330nm) bands to minimise transmission losses in optical fibers, in the visible range to stablish optical links in free space, or at 780/795nm to be compatible with quantum memories based on rubidium vapour [13]. The emission may also need to be polarised, emit with a desired spatial distribution or with a given orbital angular momentum, or emit with a controlled temporal shape.

Different emitters and materials can give rise to a **variety of energy level structures**, some of which may prohibit certain applications, even if most other criteria are satisfied. This structure also significantly impacts the emission dynamics, which governs the quality and shape (in space and energy) of emitted single photons [14]. As a result, certain emitters may allow for operation techniques that can greatly enhance the quality of emission over others [15–17].

Importantly, as we will discuss later in this review, most quantum applications require emitters that are effectively isolated from their environment during the emission timescale, which is an inherently difficult task to accomplish for solid-state sources that worsens with increasing temperatures. Although it is unlikely that this general trend can be completely overcome, the search for emitters that exhibit a reduced interaction with their host material promises to **increase operation temperatures**, thus decreasing the complexity of quantum photonic devices, increasing their reliability, and reducing their energy footprint [18].

Furthermore, single-photon sources constructed from solid-state emitters have the potential to contain a local quantum memory or memory register in the form of a particle **spin or nearby nuclear spin states**. This feature can unlock a wide range of additional applications that are unavailable to other source types. For example, a memory can be used to deterministically generate entangled photonic states [19,20] to implement measurement-based quantum computing [21], fusion-based quantum computing [22], or all-photonic quantum repeaters [23]. The memory states may serve as one or more physical qubits in hybrid spin-photonic quantum information architectures [24,25] or to distil entanglement in quantum networks [26]. Some emitters contain spin states that can be monitored optically even at room temperature, allowing the emitter to function as nanoscale magnetometers [27] for biological applications. Thus, *the search of new emitters brings along the exciting possibility to uncover an emitter that will disrupt the current quantum technology landscape*.

Another crucial aspect in the research of quantum light sources is technology **reproducibility and scalability**: this is a barrier that separates fundamental research from real-world applications. So far, it is difficult to determine the best options to industrialise the production of single-photon sources in (solid-state) platforms at a large scale, while maintaining high quality and reproducible performance (such as spectral emission, temporal characteristics of the single photon pulse, etc). This aim is very present in most of research groups and start-ups in the quantum industry ecosystem.

All these many factors give rise to a pronounced **trade-off among solid-state emitter platforms**, depending on the application. Currently, there is no single platform that can satisfy all criteria for an ideal single-photon source for every application (including its industrial scalability). This motivates the continued discovery and development of novel materials to produce sources of quantum light.

**The review is divided in a first introductory part on the fundaments of single photon emission in the solid-state** (Secs. 2-5, maybe experts on the topic could find convenient to skip this content)**, and a second part describing the recent advances in single-photon emission from three novel material platforms** (Sec. 6): emitters in transition metal dichalcogenide monolayers (TMDs), defects in hexagonal boron nitride (hBN) and perovskite QDs (PQDs). In the introductory part, we present the general benchmarking procedures for single-photon sources to quantitatively describe their performance (Sec. 2); then, we provide a description of probabilistic (Sec. 2.1) and deterministic single-photon sources in the solid-state (Sec. 2.2). Focusing on solid-state emitter platforms, in Sec. 3.1 we describe the most relevant excitation schemes for these sources, some of them profusely used in the single-photon characterisation of quantum materials and very relevant to deliver optimal control of the quantum light emission. In the next Sec. 3.2 we emphasize the relevance of the emitter coherence time $T_2$ and highlight the corresponding main sources of decoherence in the solid-state (Sec. 3.3). Section 4.1 explains the importance of light-matter coupling in optical cavities, and we present several, relevant regimes of interaction, setting didactic examples relevant to the solid-state materials of interest in this review. In Sec. 4.2 we explain the cavity parameters that condition the single-photon source brightness, discussing the main state-of-the-art works in different systems. Then, the review follows with Sec. 5.1, describing the state-of-the-art platform for single-photon generation (considering both probabilistic and deterministic sources), these are self-assembled III-V quantum dots (QDs) coupled to optical cavities; Sec. 5.2 follows giving a recent snapshot on other relevant solid-state platforms (such as erbium dopants, molecules, nitrogen-vacancy centers, etc).

After these introductory sections, the review focuses on three mentioned material platforms. The discovery of single photon emission from these materials is, in all cases, relatively new (less than 10 years) and for each material we give a common description regarding: (1) the material structure, (2) the origins and performance status in terms of brightness, single photon purity and indistinguishability (see next section), (3) the specific advantages of the material and (4) recent quantum applications. We finalise the review with a concluding section where we summarise the main achievements in these platforms and, from here, we project the potential research directions that will be explored with these disruptive materials.

## 2. Benchmarking single-photon sources

The performance of a single-photon source is benchmarked by its capacity to generate pulses of single photons at the "*push of a button*", one-by-one, and such that all emitted photons are identical in all relevant degrees of freedom (polarisation, spectral shape, temporal mode shape, spatial mode shape) [28]. These three criteria are quantified by the brightness ($B$), the single-photon purity ($P$), and the indistinguishability ($I$), respectively. If these **BPI values reach unity, the single-photon source is optimal** for quantum communication and computation applications (with some applications requiring only a subset of the three criteria, for example, some quantum key distribution protocols).

The brightness $B$ quantifies the probability to produce one photon with each excitation pulse. It is susceptible to inefficiencies in source generation, photon collection, and detection. Thus, the value can depend substantially on where in the optical setup $B$ is measured; in Fig. 1a, as an example, we indicate three brightness points defined in different parts of the setup from source to detector: at the first collection element (usually called first lens brightness, $B_{fl}$), coupled inside a single mode fiber ($B_f$), and at the detector ($B_d$). Along the review, $B$ will refer to the brightness at the source level, before the first lens, unless indicated otherwise with the corresponding subindex.

The single-photon purity $P$ is related to the probability that the source produces at most one photon. It is a loss-independent quantity that we define in this review as $P = 1-g^{(2)}(0)$ (following the definition of other works), where $g^{(2)}(0)$ is the integrated second order intensity correlation function at zero delay measured using a Hanbury–Brown and Twiss setup [29], see scheme of such setup in Fig. 1b. We would like to remark that here the term "single-photon purity" is not related to the purity of a quantum state, but to the degree of antibunching in the $g^{(2)}$ measurement (from here onwards in the review, $g^{(2)}(0)$ will simply write as $g^{(2)}$); we follow this nomenclature for the sake of simplicity as it is now the most extended one in the community. However, we recognise that such a use of "purity" could induce some confusion in quantum optics. Other recent methods to benchmark the capacity of a source to emit photons "one-by-one" (single-photon purity) are based on other tools beyond $g^{(2)}$, such as the P-distribution non-positivity or the non-Gaussianity of the emitted state [30].

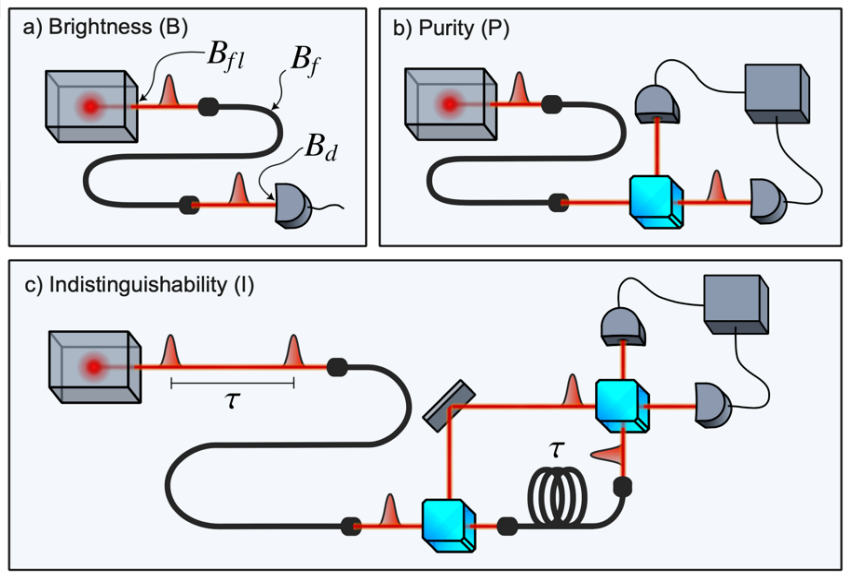

**Figure 1. Benchmarking the single-photon performance of a solid-state source.** (a) Brightness characterised under pulsed excitation conditions and defined at a given position of the collection setup. (b) Single-photon purity characterised with the integrated second order correlation function ($g^{(2)}(0)$) in a Hanbury-Brown & Twiss experiment: the level of single photon purity is P=1-$g^{(2)}(0)$. (c) Indistinguishability, measured in a path-unbalanced Mach-Zehnder interferometer, where now the degree of indistinguishability can be assessed by $I \sim 1-2g^{(2)}_{HOM}(0)$ for perfect single photon purity.

Lastly, the indistinguishability $I$, also a loss-independent quantity, captures the degree to which two photons (typically from the same source) will interfere with each other in a Hong-Ou-Mandel (HOM) setup [31]. This fundamental quantum property is typically measured in solid-state single-photon sources with the experimental setup sketched in Fig. 1c.

While a few quantum communication protocols only require high $B$ and $P$ (one renown example of this the Bennet-Brassard '84 (BB84) protocol [32]), most applications in quantum computing also demand a high degree of $I$. $B$ and $I$ are crucial for applications exploiting the interference of many photons. This is because the probability $\sim B^N$ to detect $N$ photons in a single realization of a protocol will decrease exponentially with decreasing $B$. Similarly, the probability that all $N$ photons successfully interfere is related to $I^N$. In the context of computing, if $B$ and $I$ are not sufficiently close to 1 for a given $N$, then the photonic experiment can be efficiently simulated using a classical computer [33,34], erasing any potential for a quantum advantage. Importantly, **all three of these properties must be characterised *under pulsed excitation conditions***, in which the source delivers a single photon wavepacket well-identified in time and at a periodic clock rate. The *BPI-benchmark* thus constitute a common reference for single photon generation across quantum photonic platforms that objectively describes the level of maturity and quantifies their technological readiness for real-world applications.

It is worth remarking that non-unity $P$ also degrades the amount of interference effectively observed and can influence the measurement of $I$, although the severity of its impact on protocols depends on losses and is generally quite complicated to predict. Because of this, it is common to decouple the contributions from $P$ and $I$ by applying a correction factor to the measured HOM visibility $V_{\text{HOM}}$. This correction depends on the exact cause of imperfect $P$, which might explain the fact that there is not yet a unified consensus on how to perform the correction. However, it has been experimentally demonstrated that the indistinguishability is bounded by $M \leq I \leq M/P$ in the limit that $P$ approaches unity and where $M = V_{\text{HOM}} + g^{(2)}$ is called the *mean wavepacket overlap* [35]. The indistinguishability also saturates at the upper bound $I = M/P$ for sources where any unintentional photon emission is completely distinguishable from the anticipated photon, which is the common scenario for sources based on classically driven single emitters.

## 2.1. Probabilistic sources of single photons

Historically, nonlinear processes such as spontaneous parametric down conversion (SPDC) or four-wave mixing (FWM) [36] are the most commonly used approaches to generate single photons due to their compatibility with integrated photonic circuits, relatively simple operation (cryogen-free), and access to multi-colour operation (such as visible/telecom range for free-space/fibre-based networks). The main constraint for these platforms is an intrinsic compromise between $B$ and $P$ [28].

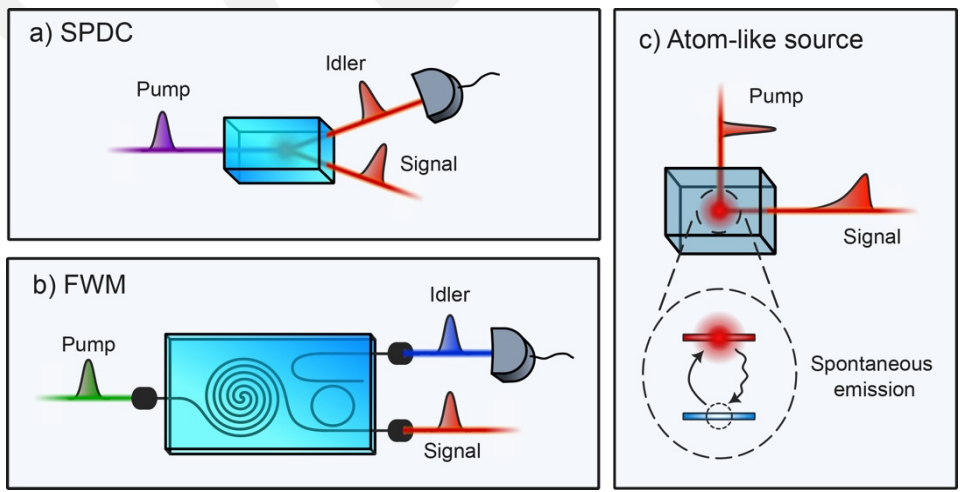

**Figure 2. Probabilistic and deterministic approaches for the generation of single photons**. (a, b) Probabilistic sources using (a) SPDC in bulk, representing as an example of a type-I conversion, one photon detection heralds the presence of another photon in the other spatial mode; (b) FWM on chip, the output generates two single photons of different energies, also photon-heralding is used to announce the presence of another single photon. (c) Atom-like solid-state

sources based on spontaneous emission from a two-level emitter: there is not heralding, a laser pulse delivers a single photon pulse in the ideal case.

These sources are called **probabilistic** because, for every pump pulse, they have a probability to produce some number of pairs of single photons. If the source produces exactly one pair, then detecting one photon of the pair called the idler will *herald* the presence of the other single photon called the signal (see Fig. 2). A simple model for the emitted quantum state of light is a two-mode squeezed vacuum state $|\psi\rangle = \sqrt{1-|\lambda|^2}\sum_{n=0}^{\infty}\lambda^n|n,n\rangle$ [37], where $|\lambda|^2 = \tanh^2(r)$ and $r$ is a dimensionless squeezing parameter that is proportional to the pumping laser amplitude. This implies that the probability of generating exactly one pair $|1,1\rangle$ used to herald a single photon increases with the pump power and ideally follows $B = \tanh^2(r)/\cosh^2(r)$. But increasing the pump power also increases the probability to generate terms of more than one pair $|n,n\rangle$ for $n > 1$. Thus, if there is photon loss or reduced photon number resolving capabilities, the detection of an idler photon may also imply the presence of more than one photon in the signal, thus reducing *P*. In the worst case, this leads to $P = 1 - 2\tanh^2(r)$. Therefore, to avoid low *P*, operation is usually restricted to lower pump powers (small $r$) which limits *B*.

Some recent works are advancing on improving efficiency and number-resolving capabilities to increase *P* for a given *B* by rejecting multi-pair events [38,39]. However, even in the ideal scenario where all multi-photon events are rejected and efficiency is perfect, the fundamental upper bound on *B* is still just ¼ achieved with a squeezing of $r = \log(\sqrt{2} - 1)$. To reach $B = 1$ using probabilistic sources, it is necessary to fabricate and synchronize multiple sources in a switching network to build an effective deterministic single-photon source. This multiplexing solution can enhance the B performance [40], however it requires substantial resources to implement, and the complexity introduces additional losses that can again degrade *B* and *P*. In contrast, solid-state emitters can surpass this photon generation efficiency with substantially fewer experimental resources [28].

## 2.2. Spontaneous emission in solid-state sources

As an alternative to probabilistic sources, solid-state-based emitters are gaining relevance for the generation of quantum light [41,42]. For these emitters, single photons are produced in much the same way as for a natural atom. Typically, an excitation laser pulse is used to deterministically prepare the emitter in a higher-energy electronic state. The relaxation of this state to a lower-energy state triggers the release of a single photon via spontaneous emission.

Once excited, the timescale for photon emission is governed by a characteristic exponential decay of the emitter excited state with a **lifetime** $T_1$, which is typically on the order of 1 ns for the three main materials discussed in this review, and typically faster than the characteristic decay time of natural atoms. The measured lifetime can be decomposed into *radiative* and *non-radiative* contributions $1/T_1 = \gamma = \gamma_r + \gamma_{nr}$. The total decay rate $\gamma$ can be determined by fitting the exponential decay of the photo-count intensity measured in a time-resolved single photon counting experiment with pulsed excitation. Accessing the value of the non-radiative decay rate $\gamma_{nr}$ separately requires experiments exploiting a configuration first introduced by Drexhage [43,44]. In this technique, the lifetime of an emitter placed closely above a mirror oscillates with emitter-mirror distance following the variations in local density of states of the electromagnetic field above the mirror. Since only $\gamma_r$ is modulated, the amplitude of variations in $T_1$ gives access to the relative magnitude of $\gamma_r$ and $\gamma_{nr}$ naturally. Suppressing the $\gamma_{nr}$ channel of decay is essential to increase the overall brightness of the solid-state emitter. We quantify the proportion of radiative decay using the quantum efficiency parameter $\eta_{QE} = \gamma_r/\gamma$, which is unity in the ideal case. Later, we present a comparison of the decay times ($1/\gamma$) among solid-state emitters and their range of quantum efficiency ($\eta_{QE}$) in Table 1.

The triggered photonic emission (see Fig. 2c) provides a protocol for the generation of quantum light that, unlike probabilistic sources, is not restricted by a fundamental trade-off between *B* and *P*, and can exceed $B = 1/4$ without a switching network [45]. A solid-state "artificial atom" can be re-excited with a pulsed laser to emit consecutive single photons with a repetition rate limited by $\gamma$, this is typically in the GHz range, but the acceleration of the spontaneous emission could bring this maximum generation rate above 10 GHz [46]. On the other hand, probabilistic sources can achieve higher rates, since the generated single photon wavepacket inherits the temporal shape of the pump laser pulse, allowing excitation rates of 50 GHz [47], but, in principle, compatible with THz regimes. However, such a high rate is not currently compatible with low-loss switching networks that are necessary to increase brightness [48], and these components will ultimately dictate the protocol rate.

## 3.1 Typical excitation schemes in solid-state sources

To exploit spontaneous emission in a two-level system, it is necessary to first prepare the emitter in an excited state. There are a variety of different excitation schemes for atom-like sources. Each scheme brings different advantages and disadvantages, and they often provide different excitation probabilities ($\eta_{exc}$) in addition to affecting other source characteristics. They can generally be classified into two groups: **resonant and non-resonant driving conditions** (see Fig. 3). In the following, for the sake of simplicity we consider excitation schemes in a two-level system, and we will not discuss other excitation schemes for the generation of entangled photons pairs (biexciton-exciton cascade), or spin-photon entanglement (schemes to control the coherent spin dynamics).

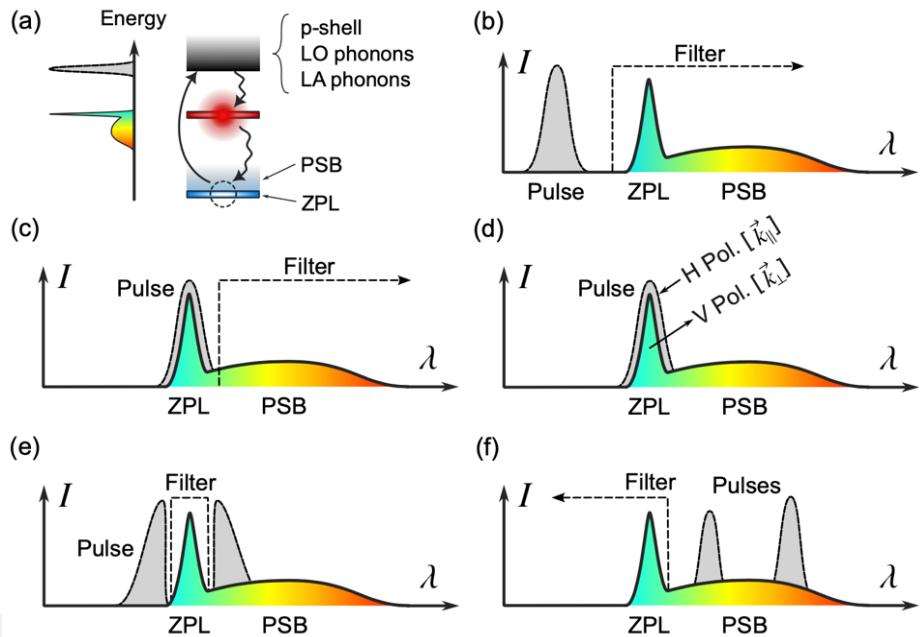

**Figure 3. Different excitation schemes in solid-state single photon sources**. (a) Energy-levels sketch of a two-level system, optically driven at a higher-energy (in schemes such as p-shell, LO, or LA phonons). (b) Corresponding excitation scheme where the driving laser and ZPL are detuned; the lowpass filter allows to spectrally filter the single photon emission. (c) Resonant drive of the ZPL, spectrally filtering both the laser and ZPL and collecting the PSB emission of single photons (common approach in hBN defects). (d) Resonant drive of the ZPL via polarisation filtering (drive and emission present orthogonal polarisations) or exciting/collecting in different directions (often implemented in photonic crystal cavities and emitters embedded in waveguides). (e,f) Bi-chromatic excitation, two pulsed lasers drive the ZPL with two colours, located surrounding the ZPL (panel e) or below it (panel f, this is the "swing-up of quantum emitter population", or SUPER scheme, see text); the laser is spectrally filtered.

In the first group of resonant driving, a pulsed laser source is tuned to match the energy of the transition that will produce the single photon (Figs. 3c,d). Then, by adjusting the excitation power (or its temporal duration), the emitter can be brought near-deterministically into its excited state at a specified time, ready to emit a single photon via spontaneous emission [49]. In this case, certain experimental schemes must be implemented to supress or filter out the laser scattered into the same spatial mode as the emitted photon. Most commonly, this is accomplished using polarisation filtering that can halve the emission brightness [50–52] (Fig. 3d), and it is generally implemented on an emitter composed by three (for example, a neutral exciton with a nonzero fine structure splitting), or four energy levels (such as a charged exciton), where the single photon emission can take place in different polarisations. Alternatively, it could also be possible to use a two-level system whose transition is, for example, horizontally polarised, and the excitation/collection is performed in the diagonal/antidiagonal polarisation basis. Other approaches can excite resonantly and collect light in different directions (for example, in emitters embedded in integrated waveguides), this is commonly done with natural atom systems [53].

As will be discussed later in Sec. 6.2 devoted to hBN-based emitters, another technique to filter single photons when using resonant excitation is to drive the zero-phonon line (ZPL) of the emitter and collect emission from the phonon sideband emission (PSB), after filtering in frequency (Fig. 3c). This technique is a suitable way to test the system response. However, it is not an optimal way to efficiently generate single photons because a significant amount of spectral filtering is required to achieve high I, which drastically reduces B. For the sake of clarity, in Fig. 3a we sketch the ZPL and PSB, together with the energy levels of the atom with the corresponding energy broadening giving rise to the PSB.

In the second group, a blue-detuned laser can prepare the emitter in a higher-energy that will then relax via photon or phonon emission processes until reaching the excited state of the desired single-photon transition (Fig. 3a,b). These non-resonant schemes are experimentally easier to implement: the laser pulse and the single photon emission can differ substantially in colour, allowing for spectral filtering to separate the quantum emission from the laser pulse in the collection path without sacrificing emission brightness. However, non-resonant schemes generally produce lower-quality single photon wavepackets compared to resonant schemes, with a lower *I* due to a reliance on dissipative processes, and possibly reduced *P* due to the increased number of recombination channels.

Equivalently, for some emitters such as QDs, a voltage bias can load a high-energy electron-hole pair that will decay and eventually lead to single-photon emission [54]. This driving scheme is equivalent to a non-resonant drive of the source and might offer some technological advantages (laser-free generation of single photons, integrability and compactness of the device), however sacrificing the optimal coherence in the single-photon emission. We will not further discuss the possibility to electrically drive solid-state sources, a review on this topic can be consulted in Ref. [55].

Excitation schemes exhibiting a trade-off between *B* and *I*, *P* is not satisfactory, since the initial promise of single emitters is to circumvent this exact trade-off experienced by probabilistic sources. Luckily, the physics of solid-state emitter dynamics is extremely rich, leading to many alternative "quasi-resonant" schemes that can harness the benefits of both resonant and non-resonant approaches. In these types of schemes, excitation is accomplished by exploiting one or more near-resonant pulses to predictably manipulate the emitter into its excited state. For example, a two-photon process can be used to target specific high-energy states of the emitter (common in the generation of entangled photon pairs in the biexciton-exciton cascade) [56]. This state can then naturally decay as in usual non-resonant schemes, or its decay can be stimulated with a second pulse [15–17] offering improved *I*. It is also possible to exploit the difference in phonon emission and absorption probabilities to harness otherwise harmful dissipative effects to excite the emitter, as in longitudinal acoustic (LA) phonon-assisted excitation [57]. Controlling interference due to multi-coloured pulses can also bring advantages: dichromatic shaping of resonant laser pulse [58,59] or near-resonant pulses [60] allows for improve excitation compatible with spectral filtering (see Fig. 3e,f). To the best of our knowledge, these schemes have only been tested in self-assembled III-V QDs, but it would be interesting to see their suitability in other solid-state emitters.

All these alternative schemes share the advantage of separating the laser pulse from the emitted single photon without polarisation filtering, and so collecting single photons in any polarisation state (essential in spin-photon interfaces). In addition to allowing unity *B* in principle, they also allow polarisation to be used as a qubit degree of freedom that can be entangled with a degree of freedom of the emitter, such as the electron spin, provided that the emitter level structure allows for it. Such spin-photon entanglement can be exploited to deterministically generate more complicated multi-photon states such as linear cluster states [61,62] and Greenberger–Horne–Zeilinger (GHZ) states [63,64] (additionally, we note that GHZ states can be generated in integrated circuits via heralding, using bright solid-state single-photon sources [65–67]).

### 3.2 Coherence time ($T_2$) in the single photon emission

The **coherence time ($T_2$)** in a solid-state emitter describes the time scale at which the phase of the two-level system is preserved during its decay process from the excited to the ground state. This coherence time is transferred to the single photon wavepacket along the spontaneous emission process, which then dictates the indistinguishability. These sources present dephasing channels that critically influence $T_2$. This effect results in a homogeneous broadening of the emitter spectrum ZPL, which is usually modelled by a Lorentzian function with a full width at half maximum of $\Gamma = \gamma + 2\gamma^* \propto 2/T_2$, where $\gamma^*$ is a phenomenological pure dephasing rate (describing the rate at which the phase of the emitter is blurred), and $\gamma$ is the total decay rate of the emitter. In the ideal case where $\gamma^* = 0$, the relation between the spontaneous emission ($T_1$) and the coherence time is $T_2 = 2T_1$. At this condition, the ZPL will display a so-called "*Fourier transform-limited linewidth*" [68], where the Fourier transform of the emitted temporal wavepacket (mono-exponentially decaying with $T_1$) will match the ZPL spectrum width ($1/T_1$). This evidences that **no phase averaging is reducing the quantum-coherent properties**, and thus the emitter should produce perfectly indistinguishable photons. In general, solid-state emitters will display a limited coherence time, implying a non-unity indistinguishability roughly given by $I = T_2/(2T_1) < 1$. A reduced temporal coherence compromises the capacity of photons to interfere, which limits the applications that depend on photon entanglement such as entanglement swapping, quantum teleportation, multi-photon interference in linear-optical gates, and photonic quantum computing.

The **measurement of $T_2$** can be performed in different ways. The "simplest" one consists of scanning the frequency of a resonant laser across the emitter transition, providing the value of $\Gamma$. The scan speed will uncover dephasing mechanisms up to the corresponding time scale of the scan [69]. Another possibility is to interferometrically measure the first order correlation function ($g^{(1)}(\tau)$) of the single photon emission versus time delay $\tau$ in a Michelson interferometer; the exponential decay of the interference visibility is dictated by the coherence time $T_2$ [70]. This option is convenient when the single-photon source is dim, since this is an *intensity measurement* that only requires a single detector. If the source is brighter, it is possible to measure *two-photon correlations* instead. By correlating the output detectors of the interference for different delays in a Michelson interferometer, the timescale of spectral fluctuations of an emitter can be retrieved with a high resolution. This is known as *photon correlation Fourier spectroscopy* [71,72], and it is notably used for the spectral characterisation dynamics of solid-state emitters [73–75].

Although the measurement of $T_2$ can give a rough idea of the indistinguishability value, the ultimate benchmark is to implement *HOM interference between two photons* [31] subsequently emitted from the same source or emitted from two remote sources. When the indistinguishability is tested for photons emitted from the same source, a path-unbalanced Mach-Zehnder interferometer is usually used (see Fig. 1c), allowing to interfere single photon wavepackets from two delayed pulses ($\Delta t$). If these single photons remain indistinguishable for very long delays [76–78], it is possible to demultiplex up to $\sim \gamma \Delta t$ photons for multi-photon applications (see Table II in Sec. 8) [65,79].

### 3.3 Pure dephasing in the single photon emission

In the following, we summarise the **main sources of dephasing** in solid-state sources. Solid-state emitters are embedded in a material matrix that opens decoherence channels due to the interaction with the surrounding atomic lattice, suffering from collective vibrations, fluctuations in the charge landscape and in the nuclear spins: these can dephase the artificial atom at various rates. The relevance of each dephasing mechanism is generally material-dependent and so, enhancing $T_2$ might require tailored solutions for each quantum material.

The impact of these environmental interactions depends on the emitter's charge state (whether it has a neutral, positive, or negative charge) and on its structural symmetry, particularly in the case of point defects. Lattice vibrations lead to changes in the emitter energy via absorption/emission of phonons, producing temperature dependent PSBs, which are offset from the emitter ZPL by the respective phonon energies (see sketch in Fig. 3a). This interaction takes place via longitudinal acoustic or optical phonons, which couple to the emitter via mechanical strain. Phonons may either be emitted (Stokes shift) resulting in a PSB red-detuned from the ZPL or absorbed (anti-Stokes shift) resulting in a blue-detuned PSB. Since the occupation of the phonon bath is dictated by a Boltzmann distribution, and the ratio of phonon emission over absorption is proportional to $(n+1)/n$ for a mode occupied by an average number of $n$ phonons, only the Stokes-shifted sideband is prominently observed at low temperatures and for phonon energies above a few meV.

The specific shape and intensity of the PSB is material-dependent and strongly depends on the microscopic spatial symmetries of the particle wavefunctions describing the emitter energy levels, which dictate the deformation-potential coupling matrix elements. The relevant parameter to consider in phonon dephasing is the comparison between light emitted into the ZPL and into the PSB; a typical figure of merit used in this case is the Debye-Waller factor, which compares the ratio of light emitted into the ZPL versus the whole (ZPL+PSB) spectrum. This factor can vary largely among different solid-state system; we compile Debye-Waller factors in Table 1 for different types of solid-state emitters.

The size of the bound electron-hole excitation (exciton) or electronic excitations in lattice defects (such as in hBN) can take several atomic lattice sites, depending on the material. Importantly, for a 3D-exciton, its size expressed via the Bohr radius is directly related to the exciton binding energy through the relation $E \sim 1/r$. This means for materials featuring only weakly bound Wannier-Mott type excitons, emission of single photons may only be expected at low temperatures, for which thermal exciton dissociation is strongly suppressed. All emerging materials discussed in this review feature comparably small excitons, which make them stable at room temperature, and – despite the mentioned impact on indistinguishability – may still be at the base of promising single-photon sources operated at cryogenic-free temperatures. For the special case of two-dimensional materials, the electron-hole pair might be in direct contact with the substrate and encapsulating material (typically hBN), leading to hyperfine energy exchange between the electron spin and nuclei of the surrounding environment (which generate a fluctuating magnetic field, acting as a dephasing channel for the emitter spin).

## 4.1 Cavity quantum electrodynamics to the rescue

Inspired by the advances on **cavity quantum electrodynamics** with natural atoms during the last decades, one special feature of solid-state sources is the possibility to engineer light-matter interactions in nano- and micro-photonic structures using optical resonators or cavities, waveguides, and topological photonic structures. This engineering provides ultimate control over the emission process by determining the mode in which the single photon emission happens in space, polarisation, energy, and time. Most importantly, coupling an emitter to an optical cavity accelerates the spontaneous emission via the Purcell effect in the weak coupling regime [80].

The presence of a cavity introduces a third decay path for the emitter. In addition to the bare radiative rate $\gamma_r$ and the non-radiative rate $\gamma_{nr}$, the emitter can also dissipate energy into the optical cavity mode via a Jaynes-Cummings type interaction with a coherent coupling rate $g$. The cavity then releases this energy into the environment with a rate given by the cavity linewidth $\kappa$. In the weak coupling regime defined by $2g \ll \kappa$, these effects combine to give the emitter a cavity-enhanced decay rate of $\gamma' = \gamma_r + \gamma_{nr} + R$ where $R = 4g^2/(\kappa + \Gamma)$ is the effective rate of emission through the cavity mode [81]. If, in addition, the cavity spectrum is much broader than the emitter spectrum such that $\kappa \gg \Gamma$, the enhancement in decay rate $\gamma' = \gamma + \gamma_r F_P$ is quantified simply by the canonical Purcell factor $F_P \equiv 4g^2/(\kappa\gamma_r)$. The cavity coupling rate $g$ and cavity decay rate $\kappa$ can be related to the cavity mode volume $V$ and cavity quality factor $Q$, respectively, indicating $F_P \propto Q/V$ is independent of the emitter properties, and depends only on the cavity design.

More generally, the emitter-cavity interaction produces dynamics that cannot be simply described by a single Purcell factor. To illustrate the various regimes of cavity coupling and the resulting quality of emission, we present the central panel of Fig. 4 inspired by Ref. [80], which shows a contour map of the product of brightness and indistinguishability ($BI$) as a function of the cavity linewidth $\kappa$ and the canonical Purcell factor $F_P$ as defined above. The panel indicates the strong coupling regime where $R < 2g$ (reached by a cavity with small volume and low losses) and the Purcell regime, part of the weak coupling regime. To best illustrate the effect of emitter dephasing, we have chosen $\gamma^* = 2.5\gamma$ when producing the plot. Such a value is similar to dephasing rates observed for TMD QDs at cryogenic temperatures [82–84]. Alternatively, the plot may represent hBN defects and PQDs at a slightly elevated temperature [73,85,86].

It is important to remark that the dephasing rate, and hence the indistinguishability, is a parameter heavily dependent on the temperature of the solid-state emitter and, in general, cryogenic temperatures are mandatory to decelerate the dephasing rates affecting the emitter. Theoretical proposals suggest specific regimes of light-matter interaction (in the weak coupling regime) that could lead to indistinguishable single photon emission at room temperature [87–90]. But achieving this might also require the discovery or synthesis of more temperature-robust emitters, perhaps in novel materials.

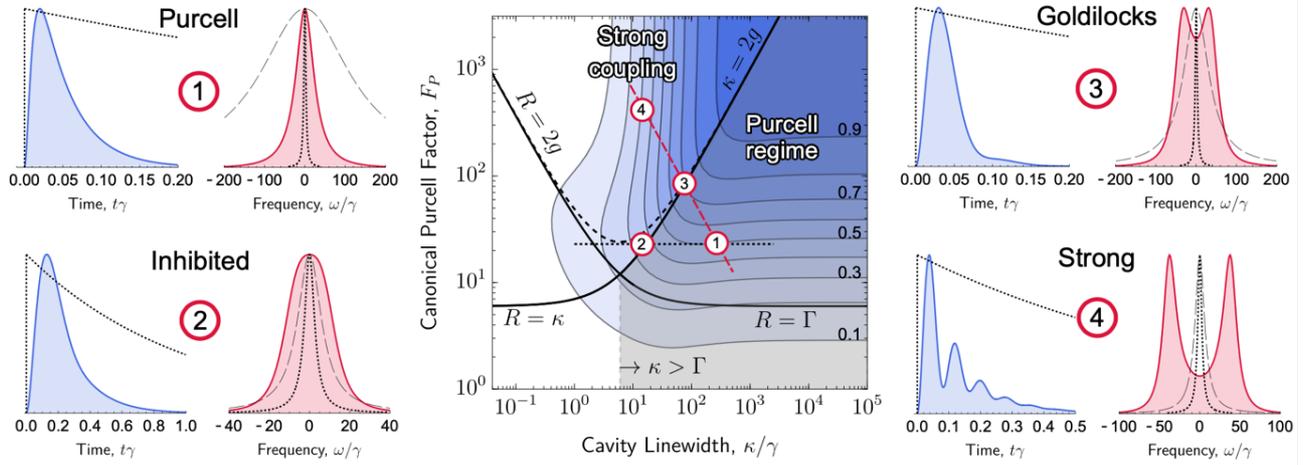

**Figure 4.** (Central panel) Product of indistinguishability and brightness ($BI$) versus cavity linewidth (horizontal axis) and canonical Purcell factor (proportional to Q/V, vertical axis). We distinguish the strong coupling regime as the region where $R < 2g$ whose boundary is illustrated by a dashed curve, and the Purcell regime as the shaded region inside weak coupling regime ($R > 2g$) where $\kappa > \Gamma$. Various features are highlighted by two thick black curves indicating different values of $R$ (see main text). The panels labelled 1 through 4 show the emitter temporal shape (blue trace) and spectrum (red trace) produced in different regimes of cavity-emitter coupling: (1) standard Purcell, (2) Inhibited Purcell regime, (3) Goldilocks condition and (4) strong coupling regime as indicated by numbered points in the central panel. The grey dashed curve in the different emitter spectra represents the cavity spectrum. The black dotted curve in the panels represent the bare decay lifetime/spectrum without a cavity. The black dotted line in the central panel indicates that points (1) and (2) share the same canonical Purcell factor. The red dashed line connecting points (1), (3) and (4) shows the slope of constant mode volume. All panels consider a pure dephasing rate of $\gamma^* = 2.5\gamma$ and non-radiative decay rate $\gamma_{nr} = 0$.

The standard **Purcell scenario** is depicted in panel (1) of Fig. 4 with $F_P = 20$, the emitter (blue trace) decays at a rate $\gamma' = F_P\gamma_r + \gamma$, ~20x faster than the no-cavity case ($\gamma$ decay rate, black dotted line). Respectively, the emitter spectrum is broadened by the cavity, with a linewidth roughly given by $\Gamma' = \gamma' + 2\gamma^*$. The high $F_P$, along with strong cavity losses, increases the proportion of light collected via the cavity mode and enhances the temporal coherence of the emission and providing $BI = 0.5$ (and $B \sim 1$) even in the presence of significant dephasing. In this regime, and neglecting other imperfections such as radiative efficiency, the improvement in brightness and indistinguishability is well-explained by $B = F_P/(F_P + 1)$ and $I = \gamma'/\Gamma'$, respectively.

The **Purcell-inhibited scenario**, depicted in panel (2), shows that the decay rate enhancement is better described directly by $R/\gamma_r$, which can be written as the canonical $F_P$ multiplied by a decoherence-induced inhibition factor $\kappa/(\kappa + \Gamma)$. If $\kappa$ is not much larger than $\Gamma$, the actual enhancement is lower than indicated by $F_P$: comparing the temporal shapes from panels (1) and (2), and noting the different x-axis scales, we can observe $F_P = 20$ relative to the case with increased cavity losses in panel (1). It is interesting to realise that the dephasing-inhibited emitter spectrum deviates significantly from a standard Lorentzian shape due to its proximity to the strong coupling regime, and that, in fact, may appear broader than both the bare cavity spectrum and bare emitter spectrum. This scenario results in relatively poor indistinguishability compared to panel (1). Yet this regime is experimentally easier to reach since it does not require as small of a mode volume and it can still provide $BI$ [80].

The "**Goldilocks condition**" depicted in panel (3), is an atypical regime on the border of strong coupling that shows behaviour qualitatively analogous to a critically damped oscillator. This condition is achieved when $\kappa = 2g$, and it provides the best possible indistinguishability for a given cavity mode volume. The emitter spectrum shows a small Rabi splitting and yet the indistinguishability is higher than in panel (1), reaching $BI = 0.7$ with the same mode volume but slightly lower quality factor.

Finally, in panel (4), we depict the typical **strongly coupled case** where the indistinguishability plummets and a clear vacuum Rabi splitting of $2g$ between two Lorentzian-shaped peaks appears in the spectrum [91]. The corresponding temporal shape also exhibits a clear beating due to the coherent exchange of energy between the cavity and the emitter.

### 4.2 Optical cavities to improve the source performance

As we will see in the rest of the review, the implementation of performant single-photon sources demands good optical cavities (high quality factor, small mode volume) in the Purcell regime [92]. These cavities can be implemented via different approaches, such as Fabry-Pérot-like cavities (with two opposing concave mirrors enclosing a resonant spacer, as seminally done with natural atoms), photonic crystal cavities, or plasmonic-like devices (such as bullseye resonators or nano-antennas where plasmonic near-fields can be tightly confined at protrusions with deep sub-wavelength radii of curvature). Different cavity approaches, with their specific properties and different regimes of operation, can be consulted in other reviews [93,42].

The photonic structure is benchmarked by the $\beta$-factor, given by $\beta = F_P/(F_P + \eta_{QE}^{-1})$ describing the probability to emit single photons in a targeted mode of the photonic structure. We recall here that $\eta_{QE}$ is the ratio between the radiative decay ($\gamma_r$) and the total decay rate of the emitter ($\gamma$). Then, single photons must be emitted with a chosen *momentum vector* allowing the photon collection, the factor representing such correct emission directionality is a ratio of photonic losses in the photonic device given by $\eta_{Coll} = \kappa_{Coll}/(\kappa_{Coll} + \kappa_{Loss})$, where $\kappa_{Coll}$ and $\kappa_{Loss}$ are the photonic losses in the desired and undesired modes, respectively. The photonic cavity is benchmarked by the product of $\beta\eta_{Coll}$, then, the total source brightness is given by $B_{fl} = \beta\eta_{Coll}\eta_{exc}\eta_{QE}$, describing the probability to retrieve a single photon pulse in the collection direction outside the source (first lens brightness) [28].

In general, the brightness can be defined at any point along the collection setup: before the first lens, at the collection fiber, at the detector, etc. Some groups quantify the brightness performance under continuous wave driving, which can be used as a preliminary study of a device. However, we think that a robust figure of merit of the source brightness must be the detected brightness under pulsed excitation $B_d = \eta_d B_f$ (where $\eta_d$ is the detector efficiency and $B_f$ would be the probability of having a single photon before the detector). **Two state-of-the-art examples from different platforms are: (1) the work of Ding et al (self-assembled III-V QDs as artificial atoms) Ref. [52], reporting a $B_d = 0.56$ with $\eta_d = 0.79$, and (2) Ourari et al (using implanted Er ions as emitters) Ref. [94], indicating a $B_d = 0.035$ with $\eta_d = 0.85$.** We choose these works as examples since they provide a full *BPI* characterisation (see Table 2 in the conclusions section).

We note that, under continuous wave excitation, one could also estimate the source brightness dividing the detected counts by the total decay rate $\gamma$ of the source (this is a common practise in emergent solid-state emitters); however, generally only a pulsed single photon stream is useful from an application standpoint.

### 5.1 A leading solid-state platform: self-assembled semiconductor quantum dots

Based on the three performance criteria defined above, over the last two decades, semiconductor QDs (coupled to optical cavities) have taken the pole position as the most efficient generators of quantum light [51,52,78,95–97]. Relevant applications in quantum communication and computation have been demonstrated (a review on this matter can be consulted, for example, in Refs. [98,99]), such as the fibre-based remote interference over hundreds of kilometres [100], hundreds of meters free-space-based [101] and kilometre-scale fibre-based metropolitan quantum key distribution demonstrations in the telecom range [102,103] (protocols where photon-interference is not required). Notably, the growth control on QDs has allowed to achieve large values of indistinguishability between photons emitted from two separate sources [104] and coupling effects among identical emitters [105–107], holding the promise to reproducibly scale-up the technology [14]. Not far behind the performance of quantum computing with cold atoms or superconducting qubits, QDs have also recently demonstrated their efficiency to generate quantum light for boson sampling experiments [79], and all-programable photonic quantum computers [65] (another review on scalable quantum photonics based on QDs can be consulted in Ref. [108]).

Recent advances have reported self-assembled QD devices interfaced with low-loss $Si_3N_4$ (~925 nm) **integrated optical circuits** [109], which are particularly important for the implementation of optical quantum processors in the wavelength range where QDs display near-optimal emission properties. The group of P. Lodahl has demonstrated single photon emission from such interfaced devices on par with the state of the art [78,110].

Beyond the single photon emission, QDs are efficient emitters of **polarisation-entangled photon pairs** via the biexciton-exciton cascade [111,112] (however with photon indistinguishability limitations [113], yet to be overcome with emitter-lifetime-engineering [15–17]), allowing the demonstration of teleportation and entanglement swapping of photonic states [114,115]. Additionally, tailoring the coupled photonic mode, the QD emission can be encoded in orbital angular momentum [116,117], suitable for information encoding in a large basis. Within the emergent field of topology in photonics we would like to mention too two works where the QD emission is released in a topological cavity [118] or topological waveguides [119]: in these approaches, information could be propagated on-chip presenting resilience against losses or defects in the material.

Overall, we believe that a good indicator for the maturity level of this technology is represented by the **number of qubits** used in quantum computation experiments [65,79] and the **distance and communication rates** in quantum communication protocols [100]. It is worth to mention as well, that numerous start-up companies have emerged exploiting self-assembled QDs for quantum optical applications; Europe hosts a big hub of them, including AegiQ (UK), g2-Zero (Spain), Sparrow Quantum (Denmark), and Quandela (France).

**5.2 More relevant solid-state platforms for quantum photonics**

The current state-of-the-art on cluster state generation based on spin-photon entanglement in QDs [62,64] has demonstrated a cluster state of up to 10 photons [61], still behind the performance attainable with natural atoms ($^{87}$Rb) coupled to cavities [120].

A technological challenge that QD sources are facing is their performance in the generation of indistinguishable telecom single-photons (mostly due to the semiconductor growth complexity in this frequency range), crucial in fibre-communication [121–123]. In this regard, other solid-state platforms, such as defect centres in silicon [124–128] and erbium ions coupled to cavities [129–132] are approaching competent levels of telecom single-photon *BPI* performance [94]. It may thus well be the case that soon these platforms will be the leading technology for optimal single-photon sources in the solid-state, compatible with industrially available silicon-based integrated photonics.

In a shorter wavelength range (~650-790 nm), suitable for free-space quantum communications, it is worth to mention the platforms of single molecules and centres in diamonds or SiC. This review will not cover the state-of-the-art of these systems in depth, but we mention that molecule-based single-photon sources (a complete review on the topic can be consulted in Ref. [133]) have shown competent degree of photon indistinguishability emitted from a single source [134,135], between two remote sources [136,137] and even the coupling between two molecules allowing to tune the sub- and superradiant nature of the emission [138]. Experiments on molecule-emitters have also demonstrated the energy-tuning of the single photon emission via strain [139] and testbed quantum key distribution schemes with molecule sources at room temperature [140].

Vacancy centres in diamond and SiC (relevant reviews on the topic could be consulted in Refs. [141–144]) are very important in spin-photon interfaces for entanglement distribution in quantum communication protocols [145], we note several recent advances on single photon generation, showing the wide tuning of the emitter energy [146], different cavity coupling implementations [147–149], nanophotonic circuits [150,151], and demonstration of indistinguishable photons from a single and remote sources [152,153]. Carbon nanotubes are also well known single photon emitters, including in the telecom range, and they have been coupled to different cavities; a review on the topic can be consulted in Ref. [154].

## 6. Emergent materials for single photon emission: TMD monolayers, hBN and perovskites

In the following, we focus the discussion on three emergent solid-state platforms for single photon generation: (1) QDs in monolayers of transition metal dichalcogenides (TMDs), (2) defects in hBN, and (3) perovskite-based QDs (PQDs). These materials present unique properties, such as the flexibility of TMDs, the room temperature operability in hBN defects and PQDs sources, and their widely tuneable emission.

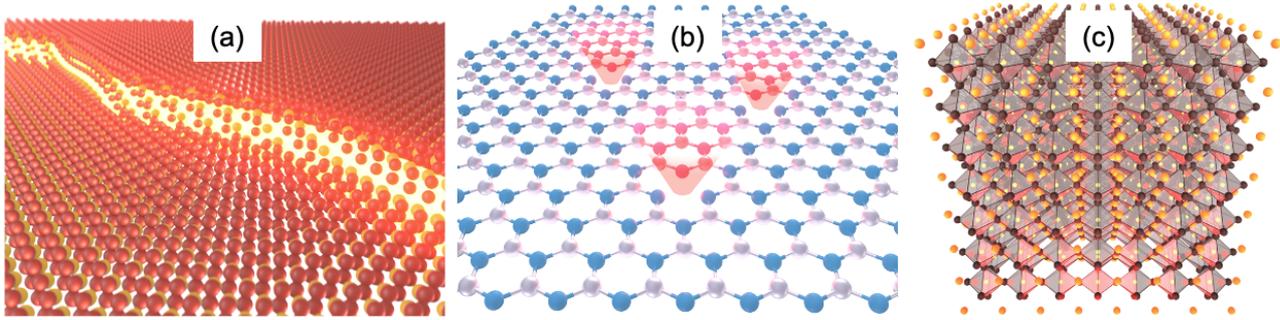

**Figure 5. Crystalline structure of the three relevant emergent quantum materials of the review.** (a) Single emitters in a monolayer of TMD (created via local strain e.g. along a fold as shown in the figure, or with the deterministic generation of a defect in the 2D crystal matrix). (b) Typical Boron vacancy defect in hBN lattice (nitrogen/boron atoms represented in blue/white colour). (c) Representation of a cubic perovskite nanocrystal.

A relevant aspect that these materials have in common is that they are **easily accessible and easily processed** for creating single-photon sources: we believe this is one of the reasons behind the large number of studies since their discovery (less than 10 years ago). None of these materials require large clean room facilities, TMDs are commercially available in their bulk form which is the basis for exfoliation and dry stamping preparation of atomically thin crystals from these van-der-Waals materials, hBN may be commercially purchased as bulk or as powder, and colloidal perovskite nanocrystals have also been commercialized and their chemical synthetisation is widely reported. Since the integration of these active materials into photonic structures does not rely on lattice matching like for self-assembled III-V QDs, they can be integrated with photonic structures (such as waveguides or optical resonators) and provide a direct path towards quantum photonic testbed applications. In addition, all three materials are relatively stable at ambient conditions. For every system, in the next sections we will discuss the state-of-the-art, the specific advantages of the material and we will review the recent applications in quantum photonics.

In the following Table 1, we compare semiconductor QDs, TMDs, hBN and PQDs attending to their different characteristics, such as their range of emission wavelength, the spontaneous emission lifetime ($T_1 = 1/\gamma$), their typical Debye-Waller factor, and the approximate quantum efficiency range ($\eta_{QE}$).

| Emitter | Wavelength (nm) | $T_1$ (ns) | Debye-Waller factor | $\eta_{QE}$ |
|---|---|---|---|---|
| Self-assembled QDs | 780-1550 | 0.3 - 1 | >0.95 (cryogenic) | 90-100% |
| TMD QDs | 600-1550 | 1 - 10000 | ~0.8 (cryogenic) | 30-90% |
| hBN defects | 450-850 | 1 - 10 | >0.70 (cryogenic) / 0.35-0.80 (RT) | 60-90% |
| PQDs | 400-800 | 0.2 - 10 | 0.5 - 0.8 (RT) | 90-100% |

**Table 1**. Intrinsic parameters from different quantum materials, such as wavelength of emission (except for PQDs, whose emission range can be continuously tuned, in the other materials we indicate the range in which ZPLs can be present at specific wavelengths), spontaneous decay lifetime ($T_1$), relevance of phonon dephasing (Debye-Waller factor, we specify if the value is measured at cryogenic temperatures or at room temperature, RT) and $\eta_{QE}$ (comparing the radiative versus the non-radiative emission rates, see main text). In the case of hBN defects, there can be many very different chemical and structural defect configurations leading to a wide range of intrinsic parameters. The range of Debye-Waller factor is estimated from about 400 different spectra measured by the group of A. Kubanek[†].

---

[†] A. Kubanek, private correspondence: Debye-Waller of 0.84±0.14 at 5K and 0.59±0.22 at room temperature from about 400 spectra of different hBN defects.

For most hBN defects, the ZPL is broadened by acoustic phonons in such a way that they cannot be spectrally filtered. In some cases, however, the acoustic phonons form a separate PSB shoulder. In this case, Debye-Waller factor is reported in some work as ~0.35 [155,156]. An example of another defect kind with phonon-broadened ZPL at cryogenic temperature can be found in Ref [157]. In this case, no proper definition of Debye-Waller factor can be given. At room temperature, the ZPL broadening by phonons is always very large. A Debye-Waller factor ~0.8 commonly found in the literature concerns the part of the spectrum that excludes optical phonon replica (i.e. the weight of the central peak) - including acoustic phonons. With this definition, III-V QDs would reach a Debye-Waller factor of 1 at cryogenic temperatures, and the two above-mentioned hBN defects kinds, with separable acoustic phonons, would also reach a Debye-Waller factor ~0.8.

## 6.1 Single-photon emitters in TMD monolayers

### 6.1.1 Structure of the material

Transition metal dichalcogenides (TMDs) are compounds with a chemical structure given by $MX_2$, where M is a transition metal (typically Mo, or W) and X is a chalcogen element (S, Se, or Te). These materials present a layered structure with hexagonally arranged X-M-X units where a layer of metal atoms (M) is sandwiched between two layers of chalcogen atoms (X); the stacked monolayers are held together via van der Waals forces. At the monolayer level, TMDs display several structural phases, commonly trigonal prismatic (2H) and octahedral (1T) phase, differentiated by the chalcogen atom positioning. In the 2H case, the chalcogen atoms from different atomic planes occupy the same in-plane positions and align vertically in the perpendicular direction of the monolayer. In the 1T case, the chalcogen atoms are placed in different positions between atomic planes (and so, vertically displaced). While bulk or few-layer TMDs exhibit layer-dependent, indirect band gaps, monolayer TMDs have direct band gaps in the range of approximately 1 to 2.5 eV [158]. This direct bandgap leads to efficient and robust photoluminescence in monolayer TMDs. Due to the strong electron-hole interaction, excitons formed in these materials possess a substantial binding energy. Single excitons can be localized or trapped by defects or local strain, providing ultimately thin single-photon emission, as we will discuss in the following.

### 6.1.2 Origins and performance status

The demonstration of single photon emission was achieved in 2015 by the set of Refs. [159–163], concerning the quantum emission from single excitons in monolayers of $WSe_2$ at cryogenic temperatures. Since then, many strategies have been implemented to improve the emission performance. Plasmonic nano-antennas of GaP [164] coupled to a single exciton in $WSe_2$ have been proven successful to generate a bright emission ($B_{ff}$~0.86), with a maximum $\eta_{QE}$~80%, and average of ~30%. The plasmonic resonance of Au nanocubes was used to decrease $T_1$ (down to ~180 ps) and generate single photons with a purity of 80% in the pulsed regime, and $B_{ff}$~0.54 [165]. These works characterise the coherence time $T_2$ (in the order of few picoseconds) via single-photon interferometry (first order correlation measurement), with values in agreement with the measured ZPL linewidth ($\Gamma$).

Other strategies to enhance the single photon emission from TMD QDs have used open cavities [166], providing a high $B_{fl}$ (65%), $P$ (>95%) and a moderate Purcell enhancement [82], which can reach up to values of 10 under proper cavity design. In Ref. [82], the high rate of collected single photons also allowed to measure $I$ via two-photon interference for the first time. Yet, the dephasing is remarkably high (on the order of few tens of ps), and only temporal post-selection of the interfered photons allowed for a non-zero visibility of photon interference. The open cavity tunability has enabled to study the cavity role in the dephasing processes of this system [84], showing that the predominant dephasing mechanisms are far from being attenuated by the current cavity characteristics. Figure 6 compiles various works on TMD QDs, reporting the *BPI* state-of-the-art values. We also note that Ref. [82] provides the first simultaneous *BPI* characterisation, with high values of $B_{fl}$ and $P$, but with values still not above the individual state of the art values described in Fig. 6.

*[Figure 6 not available]*

**Figure 6. BPI performance for single-photon sources based in TMD QDs.** (a) Brightness at the detection of $B_d = 0.024$, corresponding to the saturation of the red trace, figure extracted from Ref. [165], reproduced with permission from Springer Nature. (b) Purity of the source $g^{(2)}$~0.036, used with permission of IOP Publishing, Ltd, from Ref. [83]; permission conveyed through Copyright Clearance Center, Inc. (c) Two photon interference between successively emitted photons from the same source with a delay of 13 ns, extracted from Ref. [82], copyright 2023, published by American Chemical Society and licensed under CC-BY 4.0.

### *6.1.3 Specific advantages of the material*

Monolayer TMDs stand out as an emergent platform for single photon emission due to several unique properties: due to their weak van der Waals binding forces between monolayers, TMDs offer rich possibilities for hetero-integration including graphene gates, piezoelectric materials, and photonic circuits. This flexibility enables a wide range of photonic TMD devices showing single photon emission.

It has been shown that single photon sources in monolayers [167,168] and hetero-bilayers [169] can be electrically driven. The authors of [167] realized a photodiode-type structure comprising a single layer of graphene followed by a thin layer of hBN and a mono-/bilayer of $WSe_2$. By applying a voltage bias of ~12 V across this stack, electrons may tunnel from the graphene into the $WSe_2$ leading to radiative recombination with a main contribution from a negatively charged delocalized excitons and additional, localized red-shifted emission from quantum dots with continuous wave $g^{(2)}$=0.29±0.08, and a corresponding lifetime of 9.4±2.8 ns.

Hetero-integration also allows for the deterministic generation of quantum dots in TMDs through strain seeding. Local strain leads to a local reduction in the optical band gap effectively forming potential traps that capture free excitons and funnel them into defect-bound states whose emission is then drastically increased. Since TMD monolayers tend to form wrinkles when locally strained, the resulting strain profiles are often highly anisotropic. The resulting anisotropic trapping potential seems to be a determining factor in the observed high degree of linear polarization in TMD QDs [82,170,171] Local static strain has been induced by positioning $WSe_2$ monolayers on photonic nanostructures such as arrays of nanoneedles [171–173], by AFM nano-indentations [169,174] or by positioning the monolayer over a few-nm sized gap between nanorods of $Si_3N_4$ [175]. The latter approach has been shown to yield control over the linear polarization of emission via tuning the in-plane orientation of anisotropic tensile strain as a function of gap width. Strain may also be actively tuned in-situ when the monolayer is deposited on top of a piezo-responsive material [176,177]. Ref. [176] showed that single photon emitters in $WSe_2$ defined along a wrinkle may be tuned in emission energy over a range as wide as 18 meV. The electric tunability of the ZPL via dc Stark effect is possible for interlayer excitons in TMD hetero-bilayers in both the ensemble [178,179] and single-emitter regime [180]. Also, controlled charging of single-photon emitters has been shown for moiré interlayer excitons in TMD heterobilayers [181–183] as well as monolayer TMDs [184].

Another important strategy to deterministically induce local strain is the formation of nanobubbles under the topmost monolayer of $WSe_2$ bulk through hydrogen implantation [185]. The largest amount of strain is then found at the rim of the bubble, upon hBN capping these bubbles are also cryo-compatible.

Several strategies exist to directly control the density and location of luminescing defects in TMD materials: Helium-ion irradiation of $MoS_2$ generates long-lived vacancy defects ($T_1$ on the microsecond scale) [186,187], for which a continuous wave $g^{(2)}$=0.23±0.04 has been reported at 5 K [186]. Similarly, electron irradiation can induce vacancy emitters in WSe2 [173,188] with microsecond valley lifetimes.

The absence of crystal inversion symmetry in combination with strong spin-orbit coupling results in opposite signs of spin-orbit splitting at +K and -K valleys, leading to an effective spin-valley coupling. This distinctive characteristic grants monolayer TMDs a momentum-valley degree of freedom which results in valley-specific optical polarization selection rules and is the basis of the field of "*valleytronics*". In the context of single photon emission, long valley lifetimes for atomistic defect-bound excitons have been reported [188], yet the addition of local strain for single-emitter generation may relax the spin-valley locking selection rules [189]. On the other hand, very promising developments into the direction of valleytronic devices with $WSe_2$ QDs were achieved by the authors of [190]. They showed that under the application of a weak magnetic field and hole doping conditions, an exciton is selectively excited in the K(-K) valley using $\sigma^+(\sigma^-)$ excitation. Once this exciton captures an excess hole, the spin-valley state of the hole is anti-correlated with the exciton. The authors argue that the initialized spin-valley state of the extra hole may be much longer lived than the ns T1 recombination life time of the exciton and show that as a function of magnetic field strength the initialization fidelity of the excess hole spin may approach unity [190]

Following the demonstration of excitons trapped in the Moiré potential of bilayer heterostructures from $MoSe_2/WSe_2$ [191], it was proposed that due to repulsive meV scale dipole-dipole interactions between charge transfer interlayer excitons in this system [192], excitons in Moiré traps should potentially act as single photon emitters. While experimental demonstrations of the effect have mostly been elusive, one publication [180] indeed reports the observation of single photon emission in $MoSe_2/WSe_2$ heterostructures with continuous wave $g^{(2)}$= 0.28±0.03 in saturation, a life time of $T_1$= 12.1±0.3 ns.

Strain-induced single photon emitters in $WSe_2$ monolayers have also been integrated with $Si_3N_4$ waveguides. Ref. [193] in particular demonstrates that under continuous wave, resonant excitation through the waveguide, they achieve a $g^{(2)}$=0.377±0.081 for photons emitted normal to the waveguide surface.

Finally, we would like to note that, in 2021, Ref. [194] demonstrated telecom single-photon emission from trapped excitons in $MoTe_2$ monolayers using strain-induced nano-pillar arrays. These emitters exhibited photon antibunching at cryogenic temperatures with $P$>94%.

### *6.1.4 Quantum applications & Perspectives*

Single-photon sources based on TMD QDs have been recently used in quantum key distribution protocols, in particular the emulation of the BB84 protocol [195]. In this work, a $WSe_2$ monolayer has been deposited on a surface consisting of silver nanoparticles, inducing wrinkles in the monolayer to create QD confinement. The cryogenic source provides a detection brightness of 0.013 at saturation (under 5 MHz non-resonant pulsed drive), and the experiments are performed with a single-photon purity of $g^{(2)}$=0.13, producing a quantum bit error rate of 0.69%. The secret key rate is characterised for different emulated losses in the communication channel, and for a loss of <4 dB, the rate is in the order of $10^{-3}$ bit/pulse.

QDs in a $WSe_2$ monolayer have shown the formation of biexcitons, and the consequent cascade to emit polarisation entangled photon pairs (showing an exciton fine structure splitting of 0.4 meV) [196].

Strain-induced single photon emitters from charge transfer excitons in $WSe_2$ bilayers have as been shown to enable quantum acoustics with TMDs. Since the interlayer charge transfer exciton couples strongly to localized phonon interlayer breathing modes, single photon emission lines in this system feature a comb of pronounced phonon replicas with ~5 meV energy separation indicating the simultaneous generation of single photons and a phonon Fock state in the breathing mode oscillator [197]. Also based on phonon side band processes, entanglement of single-photons and chiral phonons has been demonstrated in $WSe_2$ [198]. Essentially, the angular momenta of an emitted single photon, i.e. its circular polarization, is entangled with the angular momentum of a chiral phonon excited during the emission process.

## 6.2 Single photon emitters in hBN defects

### *6.2.1 Structure of the material*

Hexagonal boron nitride (hBN) is composed of boron and nitrogen atoms arranged in a hexagonal pattern with covalent bonds within the single layers. In the bulk, these are held together by van der Waals forces, presenting a large bandgap (~6 eV) that renders hBN an atomically flat, high-quality insulator with strong chemical stability, widely used for two-dimensional dielectric encapsulation with TMDs. The hBN monolayer presents a direct bandgap suitable for the formation of many point defects, which can serve as robust sources of single photons that are quickly becoming a relevant spin-photon platform at room temperature. We remark that multilayer hBN presents an indirect bandgap, and the vast majority of works on hBN-based single-photon sources concern defects in multilayer hBN, which is not detrimental to the formation of defects due to their strong confinement ~2-3 Å.

The different defects found in hBN are, for example, the nitrogen vacancy ($V_N$), boron vacancy ($V_B$), anti-site carbon vacancy ($V_NC_B$), and anti-site nitrogen vacancy ($V_NN_B$) [199]. In addition, emission lines could be explained by configurations of substitutional carbon clusters [200] and/or by a doner-acceptor pair mechanism [201]. In general, the difficulty in studying and diagnosing hBN defects for practical applications is exacerbated by the fact that their structure and chemical composition are not yet well-understood. This has motivated numerous first-principles studies to uncover unique optical properties that could be used to identify specific defects [202–208].

These defects are present in different structural formats, such as nanocrystals, powder, nanotubes, epitaxial-, and exfoliated-layers. Implantation of defects can also be performed via annealing, irradiation with electrons, ions or femtosecond pulses [209,210,156,211]. The ZPL energy can appear between ultraviolet [212] to near-infrared range [210], depending on the defect formation mechanism (irradiation enables the production of deeper defects at higher energies). Concerning experimentally observed defects, several different microscopic structures coexist. Only the near-infrared range $V_B$- is well established. It is also admitted that the 4.1 eV defect [212] (likely carbon doublet [213]) and the 2.84 eV (B-centre) are two well identified families. For most others, belonging to one or more families is yet to be clarified. Similar as it happens with TMDs QDs, the irradiation has the advantage to deterministically position defects in hBN [156,211], and it is also possible to induce these emitters via local strain with nanopillars [214,215], micro-spheres [216] or nano-indentation for deterministic positioning of defects [217]. One remaining challenge (applying to all solid-state emitters) is to achieve a deterministic control on the ZPL energy and linewidth; the ZPL energy can be controlled with external tuning of a voltage bias (dc Stark effect) [218–222] or via local strain [223]. The challenge to create spectrally identical emitters with deterministic spatial positions remains open.

### *6.2.2 Origins and performance status*

The demonstration of room temperature single photon emission from hBN defects was recently achieved in 2016 by Tran et al [224,225]. Experiments on the quantum efficiency of this material (see Table 1) have shown an average value of $\eta_{QE}$~60% [226], crucial to achieve a bright emission performance. Apart from having a high $\eta_{QE}$, it is important to collect the single photon emission in a controlled mode, such as a waveguide, or a given direction of emission in free space with reduced numerical aperture: for example, in Ref. [227] up to 46% of the single photon emission is coupled into a monolithic waveguide mode. Nanophotonic structures enable the optimal collection of single photon emission; recent examples of this in hBN defects are 3D-printed elliptical micro-lenses [228], providing a collected mode with a reduced angular aperture (~4 degrees), or tapered fibers, allowing to achieve a collection efficiency of ~10% [229]. Figure 7 compiles the works in hBN defects reporting the *BPI* state-of-the-art values, which we will discuss in the following text.

Further control and enhancement of the single photon emission is achieved with photonic resonators weakly coupled to hBN defects. Plasmonic resonances in nanoparticles (of Ag or Au) have proven to reduce the emission lifetime and increase the collected signal intensity [230,231]. Also exploiting plasmonic resonances, other experiments have designed nano-antennas [232] and metal-dielectric antennas, reaching in this case a first lens brightness of 0.13 [233]. Bullseye grating cavities have also shown a significant increase of the collected intensity in hBN defects (in blue and near infrared regions of emission) [234,235]. Optical cavities for hBN emitters have been implemented in the configuration of photonic crystal cavities [236,237], demonstrating a 10-fold increase in the ZPL intensity and a weak acceleration of the total spontaneous decay time [237], and circular micro-resonators [238]. Finally, on Fabry-Perot resonators, the first implementation was achieved by Vogl et al, demonstrating a quality factor of ~2500 and a lifetime reduction factor of ~2 [239]. Another recent experiment in an open cavity configuration (with one of the mirrors attached to the collection fiber) has achieved a x50 ZPL enhancement and a significant reduction of $\Gamma$ [240]. Given the spectral width of hBN defects, in most of these cases, the device operates in a regime where $\Gamma > \kappa$, where the dephasing-induced inhibition of the Purcell factor must be considered.

A review on: "Solid-state single-photon sources: recent advances for novel quantum materials"From all these photonic approaches, the **brightest** device reported under pulsed excitation is Ref. [233], showing a detection brightness of $B_d = 0.013$ (probability of detecting a single photon), seemingly entering the saturation regime (see Fig. 7a). Other works (see for example Refs. [211,241,242]) perform this characterisation under continuous wave driving (where clear saturation curves are attainable), obtaining brightness values at the detector in the order of $B_d = 0.01 - 0.05$, nevertheless describing a single photon delocalised in time, and so, in practice, difficult to harness in quantum applications.

*[Figure 7 not available]*

**Figure 7. *BPI* performance for single-photon sources based in hBN defects**. (a) Brightness at the detection of $B_d = 0.013$ (beginning of saturation in the purple dots), adapted with permission from Ref. [233] Copyright 2019. American Chemical Society. (b) Purity of the source g(2)~0.02 under non-resonant excitation, adapted from Ref. [243], the work is licensed under the Creative Commons Attribution 4.0 International. (c) Two photon interference between successively emitted photons from the same source with a delay of 12.5 ns, reprinted figure with permission from Ref. [244], Copyright 2023 by the American Physical Society.

The record in single-photon **purity** achieved under pulsed excitation is, so far, $g^{(2)}$~0.0064 [243], which is an excellent value for an emitter under non-resonant drive (see Fig. 7b). We will not discuss single-photon purity under continuous wave drive (even though there are extensive results with very high purities, see for example Ref. [245]), since it corresponds to a single-photon generation regime with fewer potential quantum applications. The quality of the single-photon purity for hBN sources has also been benchmarked by means of the Mandel Q parameter, providing valuable information on the intensity stability of the source [246].

On the **indistinguishability** benchmark, recent experiments from the group of A. Delteil have implemented pulsed, non-resonant excitation on B centres to measure the two-photon interference (as depicted in in Fig. 1c) and so characterise the indistinguishability between delayed photon pulses [244] (see Fig. 7c). They find that successively emitted photons, with a delay of 12.5 ns, display an indistinguishability of $I$~0.56 ($T_2$~1.5 ns, $T_1$=1.9 ns), corrected by the single-photon purity $g^{(2)}$~0.14. For the sake of completeness, a recent a review on coherent single photon emission from hBN defects can be consulted in Ref. [247].

Besides the above-mentioned results, most of works to measure $T_2$ have concentrated on continuous-wave single-detector measurements such as resonant laser scans and $g^{(1)}$ (first order correlation) measurements. For the resonant spectroscopy scans, the ZPL of hBN defects is resonantly driven, detecting the incoherent light emitted via the redshifted PSB. This technique can be used to determine the ZPL linewidth (Γ) produced by fluctuation within the laser-scan timescale [248].

By resonantly scanning the ZPL with a laser, the group of A. Kubanek has observed Fourier-transform limited lines [249] ($T_2$~$2T_1$, with $T_1$=2.65 ns) up to room temperature within the ~10 ms scanning timescale. They also observed significant spectral diffusion on a timescale longer than 10 ms, which must be controlled for practical applications. A follow-up experiment gave evidence that an out-of-plane distortion can decouple the emitter from low-energy phonons, an effect which also persists up to room temperature [250]. However, recent density functional theory (DFT) simulations suggest that an out-of-plane distortion is not sufficient to provide hBN defects with Fourier-Transform limited lines at room temperature [251], and so the exact decoupling phenomenon may be more complicated. Other experiments show substantial homogeneous broadening of hBN defect lines at room temperature [85,86], indicating that Fourier Transform-limited lines are not a common feature of hBN defects.

With this same ZPL resonant driving scheme, measuring the $g^{(2)}$ histogram of the PSB photons can reveal the Rabi oscillations of the two-level system continuously driven, allowing the extraction of $T_2$ [252]. From this kind of measurement, the group of A. Delteil has implemented a recent cryogenic study showing the emission of indistinguishable photons ($T_2$~$2T_1$, $T_1$=1.8 ns) within a ~10 µs timescale [253]. It is important to mention that this excitation mechanism is a powerful tool to characterise the coherence ($T_2$) of the single photon emission, but it is not useful to generate indistinguishable single photons for two reasons: (1) the single photon generation is not in the pulsed regime, and (2) the resonantly generated photons cannot be filtered out from the driving laser field.

At this point, a work that fully benchmarks the performance of an hBN-based single-photon source (*BPI* under pulsed drive) seems to be missing, impeding a direct comparison to the state-of-the-art in other solid-state platforms.

### 6.2.3 Specific advantages of the material

Most remarkably, defects in hBN are a suitable platform for spin-photon interfacing [254]. Most of the experimental implementations are dedicated to the study of spin ensembles, which have applications in quantum sensing [255]; recent experiments from Ramsay et al. [256] have shown spin-ensembles with a spin coherence time of $T_2$=4 µs, and electron spin lifetime of $T_1$=10 µs, comparably close to the record value $T_2 \sim T_1/2$ reported in NV centres in diamond [257]. Other techniques, such as the pulsed dynamical decoupling used in Ref. [258], have extended the spin coherence time to 7.5 µs, very close to the fundamental $T_1$ limit. On single-spin detection, the first room-temperature optically detected magnetic resonance from single carbon-related defects in hBN was reported in Ref. [259]. On single-spin control, experiments from Guo et al [260] have shown room temperature $T_2$ ($T_1$) values of 2.45 µs (16 µs), controlling the spin in the hBN defect with a gold microwave waveguide. Another recent work from R. N. Patel et al [261] measure characteristic single-spin times of $T_1 \sim 100$ µs and $T_2^* \sim 6.3$ ns by making use of photon emission correlation spectroscopy [262]. It is also worth noting that the development of optically active spin defects in hBN, such as the negatively charged boron-vacancy centre, has already led to the manipulation of local nuclear spins [263]. The presence of additional hyperfine levels promises additional qubit degrees of freedom, opening the door to more complex quantum communication and information processing protocols.

One important feature of hBN defects is the possibility to deterministically implant defects in controlled locations, enabling reproducible emission wavelengths [156], which is crucial for the scalability promise of the platform. A recent review in Ref. [264] compiles the different techniques used to deterministically locate defects, and so we will not enter into significant detail. It is worth mentioning here that some of the most common techniques are irradiation with femtosecond laser pulses [211], ions [265,266], or electrons [267], and chemical vapor deposition on nanostructures [215]. Another relevant feature in solid-state emitters is the capacity to tune the ZPL energy on demand. Two main routes are generally followed here: engineered strain on the emitter or a DC Stark effect to shift the energy levels of the emitter [223,268,222].

### 6.2.4 Quantum applications & Perspectives

Given the advantage of room temperature operation, hBN sources have been used for quantum random number generation [269], a convenient application that does not require indistinguishable photons. For this, the single photon emission (under continuous wave drive) was fed into an integrated optical circuit consisting of a 4-mode splitter with outputs coupled to 4 avalanche photodiodes. The random pattern of detection clicks was then used to generate a string of random numbers benchmarked using NIST standards.

Sources based on hBN defects have also been benchmarked for quantum key distribution (QKD), which is another convenient application that does not require indistinguishable photons [270]. The group of S. Ates [271] has implemented the Bennet '92 protocol using a source without collection enhancement that provides a brightness of ~0.03 after the first collection lens and a single-photon purity of $P$=0.95. With the source driven at 1 MHz repetition rate, this QKD test achieves a secret key rate of 238 bits/s (only limited by the modulation speed of the electro-optical components that encode information in the single photon stream) and a quantum bit error rate of 8.95%. The group of I. Aharonovich has also recently implemented a QKD protocol in an underwater channel, benefitting from the wavelength of emission (~450 nm) from B-centers in hBN [272], suitable for longer transmission distances in water.

Other groups have developed hBN sources in highly compact units that can be loaded onto moving systems (satellites, vehicles, planes, etc) for free-space QKD protocols [205, 207]. In these experiments, the collection efficiency was enhanced with Fabry-Pérot cavities and solid-immersion lens systems, respectively. Moving towards the implementation of satellite based QKD, a recent work has studied the (gamma) radiation effect on hBN sources and $WS_2$ QDs, observing the resilience of hBN and a degradation in performance for $WS_2$ QDs. Along this same direction, an inspiring work directed by T. Vogl proposes a compact design for a satellite unit containing a single-photon source based on hBN emitters that will be used for QKD in space. This design is currently being integrated into a 3U CubeSat, with a launch scheduled in 2024 [273].

We want to highlight that hBN defects present very relevant applications in quantum sensing (magnetometry) that we will not discuss here, since this application is not directly related to the performance of the material as a single-photon emitter.

**6.3 Single photon emitters in perovskites**

*6.3.1 Structure of the material*

Bulk perovskites are a large group of materials with the stoichiometry $ABX_3$ (A,B cations, X anions) with the original term referring to the mineral $CaTiO_3$. Fig. 5(c) shows the ideal cubic lattice structure with the usually larger A cation occupying the voids between octrahedra formed by the smaller B cation in 6-fold coordination with the X anion. Depending on the size of A, the lattice may be deformed reducing the symmetry to an orthorhombic geometry. While chemically synthesized lead-halide nanocubes with sizes in the range from 5-15nm are the most prevalent candidates for the realization of perovskite-based single photon emitters, other important perovskite phases exist, most notably layered two-dimensional [274] perovskites with organic spacers separating individual layers. The mostly ionic character of the bonds in perovskites is a main reason for the high crystallinity of perovskite nanocrystals synthesized at moderate temperatures.

In 2023, the Nobel Prize in chemistry was awarded to A. Ekimov, L. E. Brus, and M. Bawendi for the discovery and synthesis of QDs. Over the past decade, the group of Prof. Bawendi has critically contributed to the development and characterization of the quantum emission from perovskite QDs including the recent report of competitive single-photon purity of 98.1% at simultaneous HOM visibility of 52% [275]. In this review, for the sake of clarity, we use the term "perovskite QDs" (PQDs) referring to **nanocrystals presenting the zero-dimensional confinement of single electronic excitations**, and so, generating single photon emission via spontaneous emission.

The most common synthesis routes used for PQDs are the classic hot injection method, ligand assisted precipitation, and spray synthesis [276,277]. Lead-halide PQDs in most cases present a direct bandgap which can be tuned over the range of ~400-800 nm mostly through composition control of the halides in the bulk phase, but also via dimensionality, strain, and size control over the nanocrystals [278] in the range of the typical exciton Bohr radius, which is on the order of 6nm [279]. These properties resulting in room-temperature compatibility together with high quantum yield close to 100% [73,280] make PQDs and related structures such as nanowires and platelets [281] a fast-emerging platform for a new generation of cavity-controlled single-photon sources and photon-pair generation. The outstanding performance of PQDs has been in part linked to the defect tolerance of the material as a result of a peculiar electronic band structure with both valence and conduction band formed by anti-bonding orbitals such that uncoordinated atoms only form shallow traps within the bands as well as a high formation energy for defects [282–284].

The band edge exciton in cubic lead-halide PQDs has a fine-structure with a dark singlet ground state ($J$=0) and a bright triplet ($J$=1) with non-degenerate energies due to exchange interaction. In the orthorhombic phase, the bright triplet is further split into three angular momentum projections [285]. While there is a still ongoing debate about the precise nature of the exciton dark state, it has been put forward that owing to a combination of spin-orbit coupling and the Rashba effect, the order of bright and dark exciton may be inverted [286–288] leading to a bright exciton ground state. On the other hand, low temperature magnetic studies in $FAPbBr_3$ [289] and $CsPbI_3$ [285] have shown that for these two examples the ground state is a dark singlet which may be an actual advantage for the generation of single-photon pairs through the biexciton-exciton cascade.

Some remaining challenges that are common to most PQD emitters are the sensitivity of metal halides to moisture limiting the range of possible applications. Photo bleaching does occur for halide PQDs under non-resonant driving on the scale of minutes, yet promising improvements in synthesis and post-growth surface modifications resulting in reported non-resonant pumping over the course of 2 h [290] may alleviate this current limitation. On the other hand, challenges related to intermittent luminescence (blinking) in PQDs seem partially solved through improved synthesis routes including post synthesis surface treatments and core-shell approaches which result in reduced defect densities [291,292]. Another important challenge is the toxicity of lead-based perovskites, which resulted in efforts to engineer lead-free alternatives [293,294]. To the best of our knowledge, no single photon emission has been demonstrated from any of these alternative materials.

### 6.3.2 Origins and performance status

Since the first demonstrations of single photon emission from $CsPbI_3$ (caesium lead iodide) and $CsPbBr_3$ (caesium lead bromide) nanocrystals at room temperature in 2015 [295,296], and shortly after at low temperatures for $CsPb(Cl/Br)_3$ [297], realizations of single-photon sources in PQDs have made tremendous progress. With the development of single photon emission also from organic-inorganic perovskites such as $FAPbBr_3$ (formamidinium lead bromide) [298,299] and $MAPbI_3$ (methylammonium lead iodide) [300] as well as many types of perovskite nanoplatelets [281], PQDs can be selected for single photon emission within an extremely wide range of frequencies covering the spectral range from 400 nm ($CsPbCl_3$) to 700 nm ($CsPbI_3$) for all-inorganic materials with the range extending to 800nm for organic-inorganic materials.

*[Figure 8 not available]*

**Figure 8. *BPI* performance for single-photon sources based in PQDs**. (a) Brightness at the detection $B_d = 0.011$ under pulsed drive and in the saturation regime, the corresponding value under continuous wave driving reported in the same work is $B_d^{cw} = 0.087$ (considering an average lifetime of 9.5 ns and a saturated count rate corrected for the imperfect anti-bunching of the source), figure reproduced with permission from Ref. [301], Copyright 2021, American Chemical Society. (b) Purity of the source ($g^{(2)}$~0.02), figure reproduced from Ref. [302], Copyright 2022, published by American Chemical Society, and licensed under CC-BY-NC-ND 4.0. (c) Two photon interference between successively emitted photons from the same source with a delay of ~1.8 ns, resulting in HOM visibility of 52%, figure reproduced from Ref. [275], reproduced with permission from Springer Nature.

Figure 8 summarizes some of the most recent developments in terms of characterizing PQD single photon emitters for the *BPI* benchmark: While there is currently, to the best of our knowledge, no individual study with a simultaneous characterization of pulsed *BPI*, these three parameters have been characterized pairwise in several studies. Figure 8(a) shows pulsed brightness measurements for a single 9.8±1.2 nm sized $CsPbI_3$ nanocrystal emitting at 687 nm with a photoluminescence lifetime of 7.5 ns and a quantum yield of 95±2% [277]. For the non-resonant pulsed excitation with 5 MHz repetition rate at 405 nm, the photoluminescence saturates around 50 nW with a count rate of 55 kHz corresponding to a detected maximum brightness of $B_d = 0.011$. The authors of [277] furthermore report a pulsed $g^{(2)}$=0.02 at simultaneous $B_d = 0.0084$ (16 nW pump power). Several other publications report very similar pulsed $g^{(2)}$ values in the range of 0.02 [275,302] (see panel (b)), which sets the current record for single photon purity from PQDs.

Only very recently, a study from the group of M. Bawendi has reported first HOM two-photon interferograms measured on individual 19.8±1.7 nm sized cubic $CsPbBr_3$ nanocrystals at 3.9 K [275] (see Fig. 8c). For a single fine structure split exciton line at ~532 nm selected via polarization filtering and in addition suppressing biexciton emission, the authors report a HOM visibility of 52% at simultaneous $g^{(2)}$=0.019 under non-resonant pulsed excitation at 509nm (80 MHz repetition rate with 2ns separated doublets). The doublet structure of the excitation results in the marked five-peaked structure of the HOM interferogram. It is noteworthy that the exciton shows a comparably very short lifetime of 225 ps, which was achieved without any photonic engineering, i.e., without taking advantage of cavity physics.

We note that while this is the first direct measurement of HOM interferences with perovskite single photon emitters, there has been an earlier study from the same group [73] based on photon-correlation Fourier spectroscopy [303,304] which yielded an estimate of $T_2/(2T_1)$~0.2 for individual $CsPbBr_3$ nanocubes.

### 6.3.3 Specific advantages of the material

PQDs offer several distinct advantages as a potential novel class of materials for the realization of technologically relevant single-photon sources: Most notably, via mixing different metal and halide components the emission can be tuned over an extremely wide range of 400-800nm with size control as an additional parameter to fine-tune the emission [281,305,306]. With comparably simple and well-controlled synthesis routes, PQDs with narrow size distributions can be readily mass produced at very low cost.

Remarkably, due to the unconventional electronic structure, PQDs feature extremely high photoluminescence quantum yield close to 100% while at the same time showing very short unenhanced lifetimes which enable appreciable coherence times of single photon emission even for non-resonant pumping in the absence of a cavity. This set of properties makes perovskites an extremely promising candidate for engineering the next generation of cavity-based single-photon sources, a field which is still in its infancy (see Sec. 6.3.4). Finally, it worth mentioning that PQDs due to their colloidal nature allow hetero-integration with other structures, for example nano-porous silica [307,308].

### 6.3.4 Quantum applications & Perspectives

The field of quantum applications with PQDs is still in its infancy with first proof of principle demonstration of cavity integration coming of age only very recently.

$CsPbBr_3$ nanoplatelets have been integrated into a plano-concave Fabry-Pérot microcavity of quality factor $Q\sim490$ composed of two distributed Bragg reflectors with the spacer layer filled with PMMA hosting the PQDs [309]. In this example, deterministic spatial alignment of the emitter to the field maximum of the confined optical cavity mode was achieved by pre-patterning the PMMA layer after identification of emitter positions through photoluminescence, followed by deposition of the second Bragg reflector on top of the PMMA. The authors report that for specific emission angles they achieve a significant suppression of spectral diffusion in the nanoplatelet emission and rationalize it as the consequence of a reduced effective transition dipole moment, i.e. a reduced local electromagnetic density of states via introduction of the cavity.

$CsPbBr_3$ nanocrystals have also been integrated with $Si_3N_4$ circular Bragg grating cavities [310] with quality factor $Q\sim788$. Under non-resonant continuous-wave excitation the authors find a 5.4-fold enhancement of detected emission intensity at saturation power compared to the case without cavity, accompanied by a reduction in lifetime from 175.1±54.6 ps to 89.9 ps, a 1.95-fold enhancement in recombination rate constant (Purcell factor) and a background-corrected $g^{(2)}$ value of 0.05 ± 0.024.

$CsPbI_3$ nanocrystals have also been incorporated into an open access plano-concave optical microcavity [311] composed of two Bragg mirrors, one of which holds the nanocrystals that were spin-coated from solution. Out of cavity, the crystals show a lifetime of 12.2 ns at room temperature with 40nm full-width-half-maximum (FWHM) of the emission spectrum, which reduces to 1 nm by coupling to the $Q\sim300$ cavity in conjunction with a single photon purity of 94%.

One particularly appealing outlook in the context of quantum applications is the observation of photon-pair emission in $CsPbI_3$ [285,312] through a biexciton-exciton cascade [313], which is seconded by the observation of biexcitons in $CsPbI_3$ [296] and $FAPbBr_3$ [298] and may become a key ingredient for photon-entanglement generation with PQDs.

Another recent development is the observation of photon bunching and super-bunching ($g^{(2)}(0) > 2$) in the fluorescence from lattices of PQDs [312,314,315] through the formation of sub- and superradiant modes, which may become relevant for quantum sensing applications with multi-photon states in the future.

### 7. Conclusions & Perspectives

The years 2015 and 2016 marked the starting point for the research on single-photon emission in TMD QDs, hBN defects and PQDs. Since then, the characterisation and understanding of these materials has grown exponentially and continues to achieve greater results at a quick pace. We will soon see how the *BPI* figures of merit described in this review (see table II below) are surpassed by novel cavity approaches and more ambitious applications in, for example, multi-photon protocols, source reproducibility and scalability, entanglement distribution and spin-photon interfaces. In this section, we discuss some of the conclusions found after comparing these three quantum materials and we provide perspectives on the potential research lines to come.

It was only in 2016 when self-assembled semiconductor QDs (coupled to micropillar cavities) provided a record *BPI* benchmark for the first time, obtaining impressive values in brightness and indistinguishability at that time [95–97]. These works established a clear path towards achieving higher brightness without sacrificing purity and indistinguishability. Just this year, seven years later, the group of J. W. Pan has achieved a ~70% brightness (before detection) along with state-of-the-art values for *I* and *P*, which allows for the creation of a train of 40 consecutive indistinguishable single photons at a rate of ~1 mHz [52]. This achievement is truly commendable. Regarding the novel materials discussed in this review, so far, only one work on TMD QDs (Ref. [82]) has completed a *BPI* characterisation on the same system (a QD coupled to an open cavity). After analysis, we are confident that the other two platforms (hBN defects and PQDs) will be able to be compared at the same level soon, and all three platforms may see improvements over the next years, like those seen for III-V QDs.

As discussed in the Sec. 3.1, the quality of emitted photons strongly depends on the excitation scheme. Research on TMD QDs, hBN defects and PQDs have scarcely explored resonant driving, probably rendering a sub-optimal performance in indistinguishability [82,244,275]. Notably, TMD QDs have been tested under several resonant schemes to generate single photons according to the scheme depicted in Fig. 3d [316,193]. To the best of our knowledge, the only resonant driving scheme tested for hBN defects is the one depicted in Fig. 3c [248,252], which cannot provide highly indistinguishable emission without significant spectral filtering of the PSB. Resonant driving of PQDs is yet to be demonstrated in any of the schemes discussed in Sec. 3.

As mentioned in Sec. 4, artificial atoms require a photonic cavity to quickly extract the single photon from the material (thanks to the Purcell enhancement) and control the directionality and shape of the emitted single photon mode (thanks to the structural design of the cavity). Among the different cavity-approaches, Fabry-Pérot-like cavities render the best *BPI*-performance, in both open [51,52,82] and closed [14,239] cavity versions. Photonic crystal cavities offer the possibility to control the emission in waveguides, and they seem a natural option to control many photons on chip for quantum computation applications. Self-assembled QDs [78,102] and Er ions [94] have also reached very competitive *BPI*-values in this integrated cavity approach, and seminal implementations in TMDs [175,193] and hBN [237,317] have been demonstrated recently, yet less performant than their versions in Fabry-Pérot cavities, as mentioned before.

On the quantum applications for these novel materials, we conclude that the lack of competent indistinguishability halts the experimentation on protocols that require photon interference (entanglement distribution); yet the moderate values of brightness in TMD QDs, and specially, in hBN defects, have triggered notable QKD tests. In particular, hBN-based single-photon sources have been recently developed in very compact formats (1U CubeSat structure) [239,243] aiming towards satellite-based QKD systems [273]. PQDs have a great potential to take part in quantum applications (see Refs. [275,287]), however, we think that this platform is not as advanced as TMDs or hBN, and so, more experiments on quantum functionalities are yet to be demonstrated.

There is an open problem with the *BPI* reproducibility in solid-state sources, which challenges the quick scalability of the technology towards more ambitious applications. The material matrix in solid-state systems hinders the generation of identical ZPLs in different samples (with identical spectral properties), due to the different dephasing mechanisms that intervene in the single photon emission. Some approaches in self-assembled QDs have shown some results on identical indistinguishability between 2-3 emitters [104], yet this is challenging to scale up to the levels required for fault tolerant quantum information processing. This problem also applies to single-photon emitters in TMDs, hBN and perovskites, even if defect-implantation techniques have shown a high degree of reproducibility to spectrally choose the ZPL energy in hBN defects [156].

Among the possible quantum applications, those requiring a spin-photon interface are thus far promising only for defects in hBN, as the other two novel platforms have yet to demonstrate the presence and optical manipulation of single electron spins. Notably, defects in hBN are at the forefront of room-temperature solid-state spin-photon research, with a potential for quantum applications comparable to the negatively charged nitrogen-vacancy centre in diamond. Like III-V semiconductor QDs, emitters in TMDs host an additional electron or hole to provide a spin degree of freedom, it may be the case in PDQs too. However, the experiments demonstration the coherently manipulate of spin states (to achieve spin-photon interfaces), has not yet been achieved.

*Final comparison in BPI performance*

Table 2 compiles a summary of the performance in single photon emission from self-assembled III-V semiconductor QDs, the current leader in quasi-deterministic single photon emission plus record purity and indistinguishability [52]. We also include a novel solid-state platform consisting of single erbium ions coupled to cavities (emitting in the telecom range) [94], and finally the best individual BPI performances of the three emergent materials discussed in this review. It is worth remarking that among TMD QDs, hBN defects and PQDs, only Ref. [82] has completed the triple benchmark showing high values of brightness and purity, and yet very low indistinguishability due to fast dephasing mechanisms in TMD QDs. We emphasize that the $B_d$ figure of merit could be unfair if the detection system is not optimal, however, we think that it allows to compare directly different experiments as it only depends on the detected count rates and the repetition rate of the driving. For the sake of completeness with also include a $B_{fl}$ column, where TMD QDs and hBN defects have shown a remarkable performance.

| Platform | Complete BPI benchmark | $B_{fl}$ (%) | $B_d$ (%) | P (%) | I (%) / $\tau\gamma$ | Coherent spin control |
|---|---|---|---|---|---|---|
| III-V QDs | Y [52] | 84.5* | 56 | 97.9 | 95.9 / ~50000 | Y [61,62,64] |
| Er ions | Y [94] | 72.1 | 3.5 | 98.2 | 80 / 19.4 | Y |
| TMD QDs | Y [82] | 65 [82] | 2.4 [165] | 96.4 [83] | <2 / 7.4 [82] | N |
| hBN defects | N | 13 [233] | 1.3 [233] | 99.4 [243] | 56 / 6.6 [244] | Y [260,261] |
| PQDs | N | N | 1.1 [277] | ~98 [318] | 56 / 7.9 [275] | N |

**Table 2**. Compilation of single-photon performance in several solid-state platforms indicating the state-of-the-art works in self-assembled III-V semiconductor QDs and implanted Er ions, in these two cases, the figures of merit $B_{fl}/B_d$ (pulsed, at the first lens and the detection level, respectively), *P* and *I* refer to the same work (see background colour in each row). *Ref. [52] reports the brightness inside the collection fibre ($B_f$=71.7%), we estimate $B_{fl}$ assuming the parameters $F_P$=18, $\eta_{Coll}$=0.939, $\eta_{exc}$=0.9 and $\eta_{QE}$=1 (see Sec. 4.2). For the other platforms discussed in the review, TMD QDs, hBN defects and PQDs, we indicate the best *BPI* values found in separate works. The indistinguishability value (certainly, measured at cryogenic temperatures) is accompanied by the delay (in units of the corresponding emitter lifetime) used in the two-photon interference (Ref. [52] does not include the lifetime of the emitter, we presume a decay rate of $\gamma$~1/50 ps$^{-1}$). The last column indicates whether the corresponding platform has observed a controllable spin degree of freedom in a coherent way allowing for applications requiring spin-photon interfaces.

In the comparative column of indistinguishability across different platforms, we show the delay used to test the two-photon interference. It is impressive that for the case of semiconductor self-assembled QDs, dephasing and spectral diffusion have been controlled such at a very high photon coalescence is obtained while temporally separating the two wavepackets by a delay where one could fit ~$10^4$ photon pulses. This gives an idea of the number of photonic qubits that could be (in principle) de-multiplexed into different spatial modes for photonic quantum information processing, while maintaining a mutual indistinguishability between all pairs of >95% [65,79].

In the last column we highlight the observation of an addressable spin degree of freedom in each platform, concerning works that have demonstrated the coherent control of spin levels inducing Rabi rotations and characterizing their fundamental times $T_2$ and $T_1$. This capability is essential for many quantum optical technologies relying on spin-photon interfaces.

In summary, we believe that there is a bright future for the investigation of quantum applications in TMDs, hBN and perovskite material platforms. TMDs are showing a great potential to generate correlated states of electrons, and there are many applications in quantum sensing, and maybe in quantum simulation, that could arise from this. Defects in hBN are application-relevant in QKD and we trust that, during the next 5 years, we will see this material taking an important role (together with SPDC) in satellite-based QKD communications. Certainly, there is a great application avenue for these emitters in magnetometry, due to their optimal spin control. Finally, we think that the integration of PQDs with nanophotonic structures is, comparably, less developed than the other two platforms, however, the works reported in this review have shown the excellent performance of this material to generate single photons, we are certain that the next coming years will show more applications based on photon interference.

Yet, the solid-state imposes challenges to control the emitter environment, such as the ultrafast (charge and spin noise) dephasing in cryogenic TMD QDs, and phonons in room-temperature hBN defects. For these challenges, a better understanding of the emitter physics is still required and improved cavity designs will probably be necessary. We hope that the community researching these emitters will soon discover scalable techniques to generate many identical sources (possibly via deterministic irradiation of the material) or reach regimes of multi-emitter couplings, simultaneously coupled to a single or several photonic modes (in an equivalent way as it is happening at the moment with cold atoms in lattices) that will provide a new level of control to generate states of light with higher complexity and functionalities.

## 8. List of symbols and acronyms

Following the order of appearance in the text:

Quantum dots (QDs)
Transition metal dichalcogenide (TMDs)
Hexagonal boron nitride (hBN)
Perovskite QDs (PQDs)
Brightness ($B$)
Brightness before the first lens ($B_{fl}$)
Brightness in a single mode fiber ($B_f$)
Brightness at the detector ($B_d$)
Single-photon purity ($P$)
Second-order correlation function at zero delay ($g^{(2)}$)
Indistinguishability ($I$)
Hong-Ou-Mandel (HOM)
Mean wavepacket overlap (M)
Spontaneous parametric down conversion (SPDC)
Four-wave mixing (FWM)
Spontaneous decay lifetime ($T_1$)
Total spontaneous decay rate ($\gamma$)
Radiative/non-radiative decay rate ($\gamma_r/\gamma_{nr}$)
Excitation probability ($\eta_{exc}$)
Zero-phonon line (ZPL)
Phonon sideband emission (PSB)
Longitudinal acoustic (LA)

Greenberger–Horne–Zeilinger (GHZ)
Coherence time ($T_2$)
Pure dephasing time ($T_2^*$)
Total dephasing rate ($\Gamma$)
Cavity-broadened total dephasing rate ($\Gamma'$)
Pure dephasing rate ($\gamma^*$)
Coherent coupling rate ($g$)
Cavity linewidth ($\kappa$)
Cavity-enhanced decay rate ($\gamma'$)
Effective emission rate through the cavity mode ($R$)
Purcell factor ($F_P$)
Cavity mode volume ($V$)
Cavity quality factor ($Q$)
Probability to emit photons in a targeted mode ($\beta$)
Ratio between the radiative decay and total decay
Rate of the emitter ($\eta_{QE}$)
Photon collection efficiency ($\eta_{Coll}$)
Photonic losses in desired/undesired modes ($\kappa_{Coll}/\kappa_{Loss}$)
Detector efficiency ($\eta_d$)
Density functional theory (DFT)
Quantum key distribution (QKD)

## 9. Acknowledgements

M.E. acknowledges funding within the QuantERA II programme that has received funding from the European Union's Horizon 2020 research and innovation programme under Grant Agreement No 101017733, and with funding organization the German ministry of education and research (BMBF) within the projects EQUAISE and TubLan Q.0. S.C.W. acknowledges support from the NSERC Canadian Graduate Scholarships (Grant No. 668347 and 677972). C.A.-S. acknowledges the support from the Comunidad de Madrid fund "Atracción de Talento, Mod. 1", ref.2020-T1/IND-19785, the project from the Ministerio de Ciencia e Innovación PID2020-113445GB-I00, the project ULTRA-BRIGHT from the Fundación Ramón Areces and the Grant "Leonardo for researchers in Physics 2023" from Fundación BBVA.

All the authors acknowledge Christian Schneider, Alexander Kubanek, Ferry Prins, Aymeric Delteil, Guillermo Muñoz-Matutano, and Mauro Brotons i Gisbert for their valuable comments about the manuscript. We thank Victor N. Mitryakhin for his great help with the edition of figure 5.